\documentclass[mathleft]{an}
\usepackage{graphicx}
\usepackage{natbib}
\usepackage{times}

\begin{document}

\Pagespan{1}{7}
\Yearpublication{2009}%
\Yearsubmission{2009}%

\title{A continuum of structure and stellar content from Virgo cluster
  early-type dwarfs to giants?}

\author{Joachim Janz\inst{1,2}\fnmsep\thanks{Fellow of the Gottlieb Daimler and Karl Benz Foundation.}
 \fnmsep\thanks{\email{jjanz@ari.uni-heidelberg.de}} \& Thorsten Lisker\inst{1}}

\titlerunning{Dwarf and giant early types}
\authorrunning{J. Janz \and T. Lisker}
\institute{
Astronomisches Rechen-Institut, Zentrum f\"ur Astronomie der Universit\"at Heidelberg, M\"onchhofstra{\ss}e 12-14, D-69120 Heidelberg, Germany \and Division of Astronomy, Department of Physical Sciences, University of Oulu, P.O. Box 3000, FIN-90014 Oulu, Finland}

\received{August 2009}

\keywords{galaxies: elliptical and lenticular, cD --- galaxies: dwarf --- galaxies: fundamental parameters --- galaxies: clusters: individual: (Virgo Cluster)}


\abstract{Based on the wealth of multiwavelength imaging data from the SDSS, we
investigate whether dwarf and giant early-type galaxies in the Virgo
cluster follow a continuum in their structural parameters and their
stellar population characteristics.
More specifically we study the relation between size and brightness for the galaxies and their color magnitude relation. In both cases, we find noticeable deviations from a simple joint behavior of dwarfs and giants.
We discuss these findings in the light of the different formation
mechanisms commonly assumed for dwarf and giant early types, thereby
taking into account the existence of several distinct early-type dwarf
subclasses.
By comparing our results to a semianalytic model of galaxy formation, we argue that the analyzed relations might be reproduced by processes that form dwarfs and giants altogether. 
The work presented here is based on \citealt{2008ApJ...689L..25J,2009ApJ...696L..102J}.
}

\maketitle

\section{Introduction}
Early-type dwarf (dE) galaxies are commonly expected to play a key role in understanding galaxy cluster evolution. Their importance is given by their abundances -- they outnumber all other galaxy types in dense cluster environments by far -- and the fact that they provide not too massive, not too dense test particles to probe processes that let the cluster environment alter the appearance of galaxies.
At the same time, dEs are predicted to form in models of a $\Lambda$CDM universe as the descendants of building blocks in hierarchical structure formation and to be in that sense close relatives to their giant counterparts, sharing a cosmological origin.
A better  understanding of dEs is therefore not only linked to our knowledge of formation and evolution of galaxy clusters but also of structure formation itself.

Once believed to be systems of simple appearance and well-defined properties, dEs were recently shown to exhibit a puzzling variety among their structures and stellar populations (see e.g. T. Lisker, this issue).
This diversity opens the door widely for different formation scenarios. And indeed there are different suggestions, for example the transformation of other galaxy types by the cluster environment via ram pressure stripping or harassment, which are partly able to explain  the appearance of dEs and also reproduce with some successes fundamental scaling relations of early-type galaxies.
But still today, it remains an open question to what extent these different processes play a role and
whether  some of the early-type dwarf galaxies share the same origin and formation mechanisms  with their more massive relatives. 

The above mentioned scaling relations have ever been an important tool not only to study galaxy properties but also to link those properties to their formation and evolution, 
and thus to answer the question.
Very well studied examples are 
the relations between surface brightness and size (``Kormendy relation", \citealt{1985ApJ...295...73K}),  between surface brightness and luminosity (e.g. \citealt{binggeli_cameron}).  
In combination with velocity dispersion, the Faber-Jackson relation \citep{1976ApJ...204..668F} and the extension\linebreak to the Fundamental Plane (\citealt{1987ApJ...313...42D}, \linebreak\citealt{1987ApJ...313...59D}) became famous. 
Every time these relations were analyzed for dwarfs and giants in conjunction, it was discussed, wheth\-er or not they show a common behavior and what causes it. Any dwarf formation scenario has to reproduce the observationally found relations. Additionally to these morphological and kinematical relations, the color magnitude relation (CMR), connecting the global parameter total brightness of a galaxy to its stellar population, was extensively studied, e.g.  \citet{1959PASP...71..106B,1992MNRAS.254..601B,1973ApJ...179..731F,1978ApJ...223..707S,1978ApJ...225..742S}; \linebreak\citet{1977ApJ...216..214V}. 
The CMR is typically explained by an increase of mean stellar
metallicity (and age; see e.g. \citealt{cmr_age}) with increasing galaxy mass as the dominant effect.
The common underlying idea is that more massive galaxies have deeper potential
wells, which can retain metal-enriched stellar ejecta more effectively
and subsequently recycle the enriched gas into new \linebreak stars
\citep{1997A&A...320...41K,1999ApJ...521...81F,2006MNRAS.370.1106G,2006MNRAS.366..717C}.
Also here it was explored whether and, if so, how much these processes shape the CMR of giants and dwarfs in a similar way.

We made use of a very homogenous data set of the early types in the Virgo cluster to investigate these questions via the scaling relation of size and brightness (which is a  \linebreak relative to the aforementioned morphological scaling  relations) and the color magnitude relation \citep{2008ApJ...689L..25J, 2009ApJ...696L..102J}.

\section{Sample Selection and Imaging Data}
\label{sec:imagingdata}
Our sample is based on the Virgo Cluster Catalog (VCC; \citealt{bst}). 
All early-type galaxies therein with a certain cluster member status and $m_B<18.0$ mag are taken into account,
which is the same magnitude limit up to which the VCC was found to be
complete.  
This translates into $M_B<-13.09$ mag with our adopted distance modulus
of m-M=31.09 mag (d=16.5 Mpc,
\citealt{2007ApJ...655..144M}).

Uncertain
classifications are treated as follows: galaxies listed as 
``S0:", ``E/S0'', ``S0/Sa'', and ``SB0/SBa'' are taken as S0, and
one S0 (VCC1902) is excluded, since it shows clear spiral arm
structure. For the dwarfs, we selected galaxies classified as
dE, dS0, and ``dE:", whereas ``dE/Im'' as well as possible irregulars
based on visual inspection are excluded \citep{lisker_etal}.
We exclude 37 galaxies for the following reasons: the Petrosian aperture (see below) could not be obtained, the objects were too strongly contaminated by the light of close neighbour objects, or the  $S/N$ in either the $u$ or the $z$ band was too low.  Our working sample thus consists of 468 galaxies.

The Sloan Digital Sky Survey (SDSS)
Data Release  Five (DR5) \citep{2007ApJS..172..634A} covers all but six
early-type dwarf galaxies of the VCC. 
Since the quality of sky level subtraction of the SDSS pipeline 
is insufficient, we use sky-subtracted images as provided by
\citet{lisker_etal}, based on a careful subtraction method.
The images were flux-calibrated and corrected for galactic extinction
\citep{1998ApJ...500..525S}.  
 
For each galaxy, we determined a ``Petrosian semimajor axis"  $a_p$ \citep{1976ApJ...209L...1P}, i.e. we
use ellipses instead of circles in the calculation of the Petrosian  
radius (see, e.g., \citealt{2004AJ....128..163L}). The total flux in the
$r$-band was measured within $2 a_p$, yielding a value for 
the half-light semimajor axis (SMA), $a_{hl,r,uncorr}$. 
This Petrosian aperture still misses some flux, which is of particular
relevance for the giant galaxies \citep{2001MNRAS.326..869T}. The
brightness and the half-light SMA were
corrected for this missing flux according to
\citet{2005AJ....130.1535G}.  Axial ratio and position angle were then
determined through an isophotal fit at  $2 a_{hl,r}$.
The
effective radius is then given by
$r_{\textit{eff}}=a_{\textit{hl,r}}\sqrt{b/a}$ with the axis ratio
$b/a$.  
Additionally we fitted S\'ersic profiles 
to the radial intensity profiles, keeping half light radius fixed and using an implementation of the nonlinear
least-squares Levenberg-Marquardt algorithm. For the fits we used
the intensities at $r/a_{\textit{hl,r}}=2^x$ with $x=-2 + j/4$ and
$j=0, \dots, 16$. 
We omitted intensities at radii $r <
2^{\prime\prime}$ in order to avoid seeing effects. 

Colors were measured within the elliptical $r$-band half-light aperture for each filter. 
Errors were estimated from the $S/N$ and calibration uncertainties (which we estimate to have a \emph{relative} effect of 0.01 mag in each band, which is smaller than the absolute values given by SDSS), as described in \citet{2008AJ....135..380L}.

 
\section{Sizes of early-type galaxies}
\subsection{Introduction}
The scaling relation of size and brightness of early types was not as widely studied as its relatives, like the Kormendy relation and the relation between brightness and effective surface brightness. Studies of the sizes, for example, \linebreak are: \citet{1992ApJ...399..462B,1993MNRAS.265..731G}; \linebreak \citet{1977ApJ...218..333K}, and for the Virgo Cluster in particular by \citet{binggeli_cameron} for dwarfs and by \citet{1993MNRAS.265.1013C}  for giants. But all of them share a similar history: 
 Early studies until the 1990's  came to the conclusion that giant and dwarf early-type galaxies show a distinct behavior in the scalings such as the relation between size and brightness. The dwarfs were seen to show less change of size with luminosity than the giants. This together with the other scaling relations was interpreted as evidence for a different origin of dwarf and giant early-type galaxies.

Towards the turn of the millenium, however, it became more widely realized that the light profile shapes of early types vary continuously with luminosity. Neither do dwarf galaxies simply follow exponential profiles, nor do all giants exhibit de Vaucouleurs profiles. Instead, all early types are well described by the generalized S\'ersic profile \citep{1963BAAA....6...41S} with different S\'ersic indices $n$ \citep{1994MNRAS.268L..11Y,2006ApJS..164..334F}. Several authors reasoned that the scaling relations naturally follow what is predicted by $n$ changing linearly with magnitude, and that all these galaxies can indeed be of the same kind (\citealt{jerjen_binggeli}; \citealt{1998A&A...333...17B,graham_guzman}; \linebreak\citealt{2005A&A...430..411G}). In \citet{2008ApJ...689L..25J} we studied the size brightness relation of early types in Virgo and analyzed it in the light of a continuous variation of profile shapes.

\subsection{Results}
In Fig.~1 (bottom panel) we present the size luminosity diagram for our sample. At
first glance the sequence from dwarf to giant early-type galaxies does
not look very continuous: the giants follow a steep relation with a
well-defined edge on the bright end of their distribution. The bunch of
dwarfs apparently lie with a larger scatter around an effective
radius of $r_{\textit{eff}}=1$ kpc, their sizes showing weak to no
dependence on luminosity.

\begin{figure}
\hfil\includegraphics[scale=.3,angle=0]{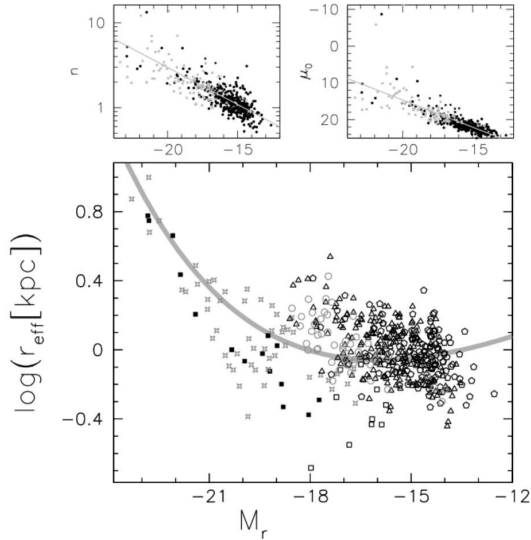}\hfill
\caption{\emph{Bottom panel:} Absolute magnitude in $r$ versus logarithm of half light radius. Filled
  squares - E, gray stars - S0, open squares - M32 candidates, open
  triangles - dE,N, open pentagons - dE,nN and gray open circles for
  dwarf galaxies with probable disk-like structure (dE(di) or dE(bc)). The grey line is calculated with the linear fits in the \emph{top panels} (see text).} 
\end{figure}

A similar impression can also be obtained from other previous studies,
e.g.\  \citealt{binggeli_cameron}, Fig.~1b;
\citealt{1992ApJ...399..462B}, Table~1; \citealt{kormendy08}, Fig.~37.
It is, however, not as clearly
seen in the compilation of sizes of elliptical galaxies from several
different studies presented by \citet{2008MNRAS.tmp..752G} (Fig.\ 10). In this
more heterogeneous data set, the relative number of small
low-luminos\-ity giants as well as that of large bright dwarfs appears to be
somewhat smaller.

\subsubsection*{Varying profile shapes ?}
\citet{graham_guzman} suggested that the apparent dichotomy between
dwarfs and giants in scaling relations can be explained just by the
fact that the profile shape of a galaxy scales with magnitude. They
describe the light profiles with S\'ersic profiles and show the effect
of a linear relation between magnitude and logarithm of the S\'ersic
index $n$ on the other scaling relations. As a result, the dependence of effective
radius on magnitude becomes stronger at higher luminosities and the brightest
galaxies are naturally larger (Fig.~11 in
\citealt{2008MNRAS.tmp..752G}).

For investigating whether our galaxies 
display the predicted behavior, 
we use the S\'ersic indices $n$ and central surface brightnesses $\mu_0$ 
to obtain linear fits to the $\mu_0/M_r$ and $n/M_r$
relations, using a least squares fitting algorithm (Fig.\ 1, top panels).  For those fits we
exclude systems with a (probable) disk component, namely galaxies
classified as S0, dEs with disk features \citep{2006AJ....132..497L},
 and dEs
with blue centers \citep{2006AJ....132.2432L}. This is to ensure that the
light profiles can be well parametrized by S\'ersic profiles.
 Our fits (Fig.~1, top panels) together with equation (16) of
\citet{2008MNRAS.tmp..752G} predict a non-linear sequence in
the $r_{\textit{eff}}/M_r$ diagram. 
The predicted relation is shown together with the observed galaxies
in the bottom panel of Fig.~1. With the visual guidance of the line,
it appears more likely that the data points follow one common continuous
relation. And the gross trend in the diagram can indeed be explained
by varying profile shapes. However, at
luminosities brightwards of the transition between dwarfs and giants, a
substantial amount of galaxies fall below the relation, while
faintwards most of the dwarfs lie above it.

As we showed in \citet{2008ApJ...689L..25J} this finding holds also if all objects with signs of disk components are omitted in order to have a purer sample of dynamically hot systems not biased by systems with more complex kinematics. Furthermore, we analyzed the deviations statistically and found significance (Fig.~2 therein). We will discuss the implications in Sect.~\ref{sec:sam} and Sect.~\ref{sec:disc}.

Our analysis showed two things. First, the distribution of data points does not
resemble a quite large random scatter around the relation. And
therefore the size luminosity relation can not be fully explained by
varying profile shapes. Second, the abrupt change in the behavior of
faint and bright galaxies is even emphasized through the above
examination,  and this break is a real
discontinuity of the sequence from lowest to highest luminosities. 

\section{Color magnitude relation}
\subsection{Introduction}

From early on, it was
discussed whether the universality of the CMR also holds over the
whole range of galaxy masses,  
i.e.~whether dwarf and giant early-type galaxies follow the same
CMR. Studies of different clusters show consistency with one
common CMR for dwarfs and giants, albeit with a
  significant increase in the scatter at low luminosities \linebreak(\citealt{1997PASP..109.1377S} for Coma, \citealt{2002AJ....123...2246C} for
\linebreak Perseus, \citealt{2003MNRAS.344..188K} and \citealt{2007A&A...463..503M} for Fornax,
\citealt{2008MNRAS.386.2311S} for Antlia, and \citealt{2008A&A...486..697M}
for Hydra I).
More explicitely, \citet{1983AJ.....88..804C} stated that there is a
common linear relation. 
But his Fig.~3 might hint at a change of 
slope from high to low luminosities, similar to what
\citet{1961ApJS....5..233D} suggested. Interestingly,
visual examination of the diagrams presented by most of the above-mentioned
studies indicates consistency also with a change of slope -- yet 
linear relations were fitted in most cases
(see, however,
\citealt{2006ApJS..164..334F}; our colors are consistent with the ones in their Virgo cluster study in the range of brightness common to both). 
In  \citet{2009ApJ...696L..102J} we revisited the question of the universality of the CMR for 
dwarfs and giants.

\subsection{Result}
We showed the CMRs for four different representative colors in \citet{2009ApJ...696L..102J}.
Here we choose to show the CMR in $u-z$, the color with the longest wavelength baseline available to SDSS.
It looks remarkably similar to the CMRs in the other colors (Fig.\ 1 in \citealt{2009ApJ...696L..102J}).

First of all, the impression one can get by examining just the \emph{black
  points} in Fig.~2 is that there is not one common linear relation from the faint to the
bright galaxies.The overall shape appears more like
  ``S'' shaped.
The brightest ($M_r<-21$) 
galaxies have almost constant color, i.e.~no correlation between color and brightness; the very brightest galaxies
show a larger scatter. These were reported before to be \emph{morphologically}
different from the other galaxies in more detailed studies of the
inner light profiles
(e.g. \citealt{2006ApJS..164..334F,kormendy08,2007ApJ...664..226L}; \linebreak\citealt{2001AJ....122..653R,2004AJ....127.1917T}). For the remaining galaxies
several descriptions seem to be plausible, ranging from just an offset
  between two relations with similar slopes up to a curved relation.

\begin{figure}
\includegraphics[scale=.4,angle=0]{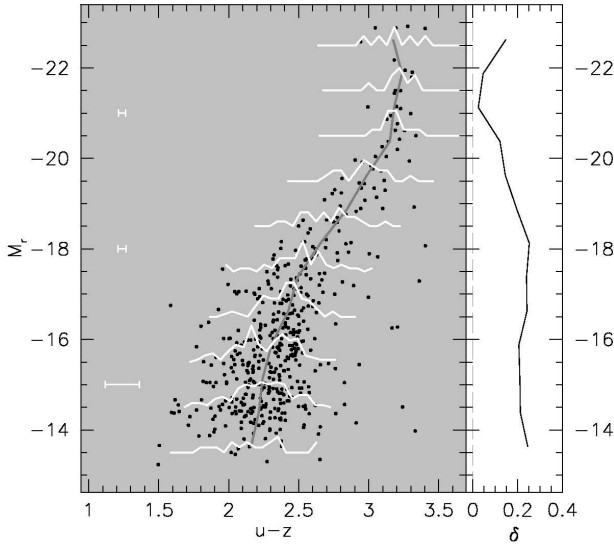}
\caption{Color Magnitude Relation in $u-z$. Filled circles show our sample's galaxies. The gray line indicates the ``running histogram'' as found
    in successive magnitude bins with a width of $1$ mag and steps of
    $0.25$ mag, clipped one time 
at $3 \sigma$.  We limit the drawing range for the
line to the region with at least three galaxies in a bin.
The white histograms show the
    distributions in bins of the same width, normalized to the square root of the number of galaxies in the bin, shown for every fourth step. 
The white errorbars indicate typical photometric errors at the respective brightness.
In the panel right
    to the color magnitude diagram a measure for the intrinsic scatter is given by the RMS difference $\delta$ of the observed scatter and the photometric error (see text) in continuous bins. }
\end{figure}

With the non-linear shape,
it seems not very favorable  to fit a straight line. This would not
describe the data well, and 
there is no theoretical prediction what other function is expected. So
at first, we want to make the overall shape more clearly visible, using
continous, overlapping magnitude bins, in which mean and scatter are calculated.
In Fig.~2 these derived relations are shown with grey lines. The first impression is
confirmed: one common linear relation for dwarfs and giants cannot be
seen. Moreover, the white histograms showing the galaxy distributions in 
the bins are clearly peaked towards the bright and the faint end,
while they are rather flat at intermediate luminosity.

The scatter about the relation is greatly influenced by  increasing photometric errors  towards faint brightness. In order to measure the \emph{intrinsic} scatter, we calculate the RMS of the scatter around the mean in
running bins (clipping one time at 3$\sigma$) and subtract  the RMS of the photometric errors 
$$\delta\equiv\textrm{RMS difference}\equiv\sqrt{\textrm{rms}^2_\textrm{scat}-\textrm{rms}^2_\textrm{err}}$$
$$=\sqrt{{\sum_i
    (c_i-\left<c\right>_i)^2}-{\sum_i \sigma_i^2}},$$
with color $c$ and mean color $\left<c\right>$, 
averaging over the galaxies in the respective bin.
Here we
exclude dEs with blue cores, since they are known to have different
colors \citep{2008AJ....135..380L}. This RMS difference should be zero if the scatter is only due to the
    measurement errors and larger if there is an intrinsic
    scatter. 
In Fig.~2 we show the CMR along with the RMS difference $\delta$.
      
Indeed, the RMS difference is enhanced
    for the dwarfs and peaks around $M_r\approx -18$, indicating an
    intrinsically increased scatter,
consistent with the increased intrinsic scatter for dwarfs found by \cite{1997PASP..109.1377S} in the Coma cluster.
 One can
    argue about the significance of the RMS difference increase for the
    brightest galaxies, since it is just a handful of them
-- nevertheless, this larger scatter might be related to the absence 
of a well-defined CMR at the brightest magnitudes.

\section{Comparison to a Semi-Analytic Model}
\label{sec:sam}

\begin{figure*}
\hfill\includegraphics[scale=.4,angle=0]{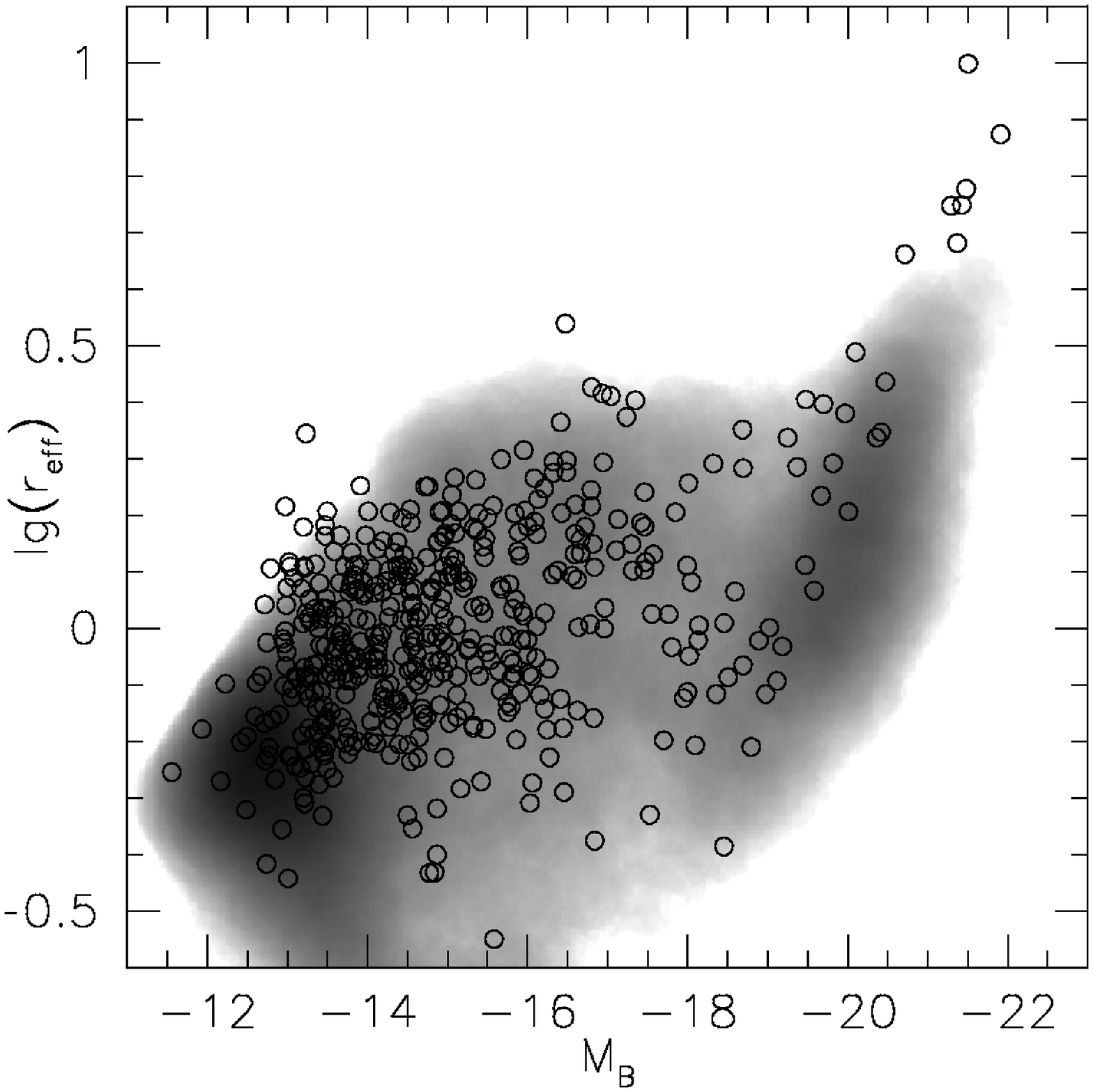}
\hfill\includegraphics[scale=.4,angle=0]{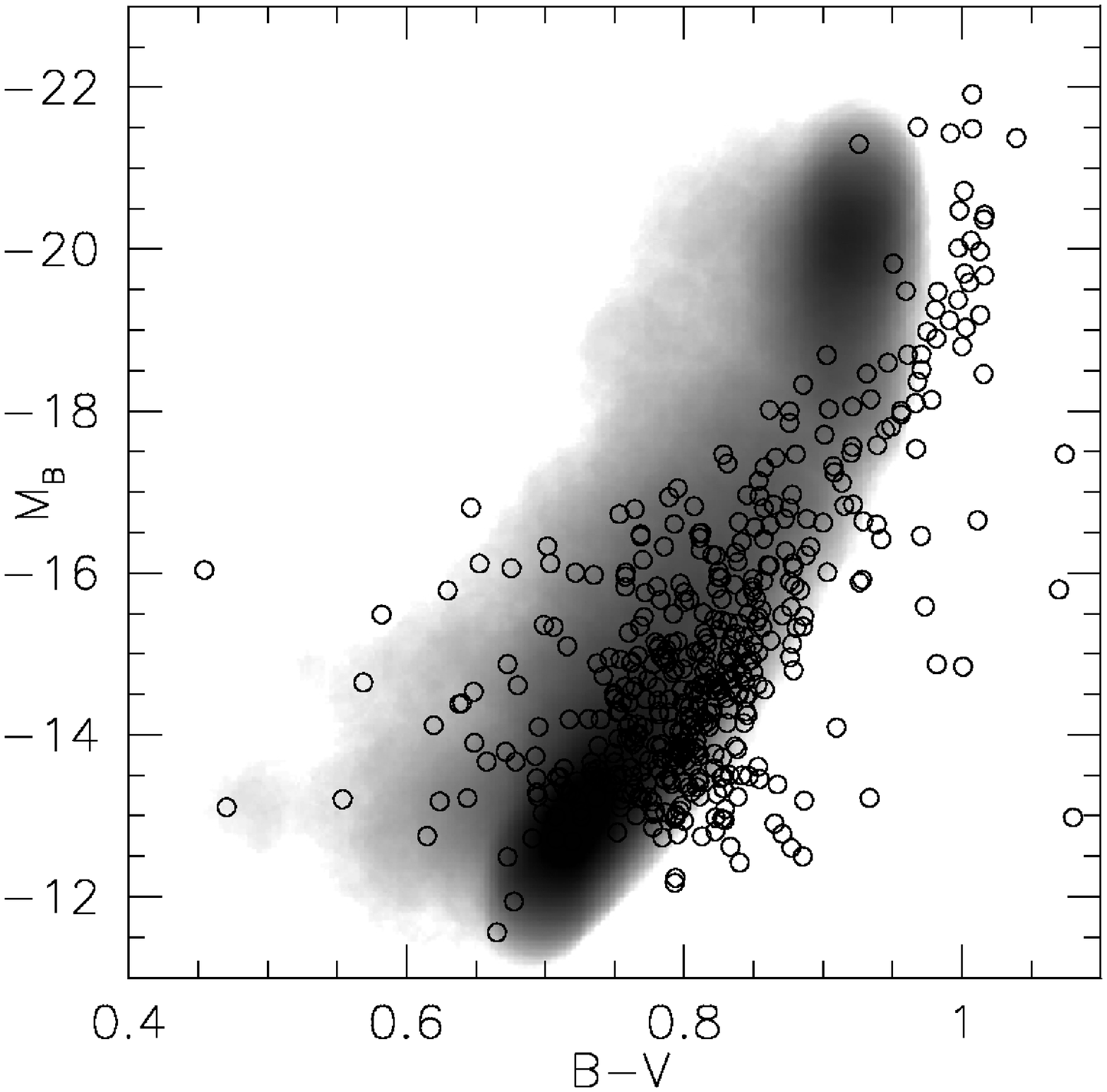}\hfill
\caption{Comparison to the semianalytic model by  \citet{2005ApJ...634...26N}. In grayscales (on a logarithmic scale) the distribution of model galaxies is shown. Model galaxies with fainter surface brightnesses than a limit of
    $\left<\mu_{B}\right><25.5\ \textrm{mag/arcsec}^2$  are excluded. The observed Virgo early type are displayed with black open circles.
        \emph{Left panel:} Absolute $B$ magnitude versus logarithm of half light radius. \emph{Right panel:}  Color Magnitude Diagram. Colors and magnitudes were transformed to the filterbands of the model output (see text). } 
\end{figure*}

Dark-halo merger trees of a high resolution $N$-body simulation of $\Lambda$CDM structure formation were taken as input for a semi-analytic model (SAM)
of the physical processes governing galaxy formation  and evolution in order to produce the Numerical Galaxy Catalog \citep{2005ApJ...634...26N}. 
In particular, the dynamical response
to starburst-induced gas removal after gas-rich mergers (also for cases intermediate
between a purely baryonic cloud and a baryonic cloud\linebreak fully supported by
surrounding dark matter as in \linebreak\citealt{1987A&A...188...13Y}) were taken into account. 
This process plays a crucial role for the sizes of early-type dwarf
galaxies,  since the subsequent variation
of the potential results in an increase in size.

We identify model galaxies as
early-type galaxies if they are bulge dominated (bulge to total ratio $>0.6)$.
To compare our data with the model we transformed SDSS $g$ magnitudes into $B$ according to \cite{2002AJ....123.2121S},
 using the galaxies' $g-r$ color measured
within $a_{hl,r}$. B-V was obtained likewise.
In Fig.~3 one can see how the model
galaxies compare with our observed Virgo early types. In the left panel the comparison is displayed for the size brightness relation. The model galaxies
 show a bimodality similar to what we observe, with low galaxy
density between the two regions. 
Note that \citeauthor{2005ApJ...634...26N} assume de Vaucouleurs
profiles to calculate projected half-light radii from half-mass radii.
For exponential profiles, which would be more appropriate for dwarfs, 
the model galaxies would  shift upwards in the diagram by 0.11 dex
\citep{2003MNRAS.340..509N}.

In the model, a starburst follows in those dwarfs that form by gas-rich major mergers  and the dwarfs are enlarged by the dynamic response to the subsequent gas loss. This mechanism is not at work in gas-deficient mergers, and the resulting galaxies stay smaller. Note, though, that this appears to be in contrast with the SAM of \citet{khochfar_sam}, where gas-rich mergers lead to more compact early-type galaxies than gas-poor mergers.

In the right panel of Fig.~3 the comparison is displayed for the CMR. The shapes of the distribution of Virgo galaxies and the model CMR are indeed not well represented by 
linear relations.
Both the CMR and the relation between size and brightness
do not show a linear, nor one common behaviour of dwarfs and giants. 
Nevertheless,  it needs to be emphasized that there is a \emph{qualitative} similarity in the shapes of observed and model CMR, in the sense that both show a similar bend at intermediate luminosities. This is noteworthy, since in the framework of the SAM, both dwarf and giants form by the same physical processes, which govern $\Lambda$CDM structure formation, and thus both can be of cosmological origin (see also \citealt{2008arXiv0812.3272C}).

Beside the similarity in the overall shape, 
an offset is observed. This offset could partly be due to
uncertainties of the adopted synthetic stellar population model. 
  Furthermore, the relative number of
bright galaxies exceeds the observed one and the luminosity function is clearly different, which could possibly be
  explained with model input physics.

\section{Discussion}
\label{sec:disc}
We studied two scaling relations of the Virgo cluster early-type
galaxies, based on model-independent size measurements from SDSS
imaging data. In both, the relation of size and brightness and the CMR, we find 
noticeable unsteadinesses between dwarfs and giants.
In the former relation dwarfs
do not fall on the extension of the  rather steep sequence of the giants.
While the gross trend in the size luminosity relation can be
explained by light profile shapes becoming steeper for more luminous
galaxies, a closer look  reveals 
a statistically significant distinctness in
the behaviour of faint and bright galaxies.
The CMR is continuous over the whole range. Yet the observed change in slope and the variation of the scatter might hint at more complex reasons for this particular behavior than what might be naively expected from having one common origin, with the very same processes shaping the CMR in the same way.

But with the comparison to the semianalytic model (not\-withstanding our rather crude way of selecting early types in the SAM) we show that neither of the two findings necessarily implies a formation by substantially distinct processes.
Instead, the qualitatively similar distributions of the model galaxies in the two scaling relations might hint at a formation in a cosmological context for the dwarf galaxies, hence an origin common to the giant ellipticals. 
This is in accordance with previous claims of no distinction be\-tween \linebreak them (\citealt{2006ApJS..164..334F,2005A&A...430..411G}; \linebreak\citealt{graham_guzman,2008A&A...486..697M}).

It must be mentioned that different approaches can also  successfully explain  the dwarf behavior. For example, ram pressure stripping can reproduce the radius brightness relation (\citealt{2008arXiv0807.3282B}, see also A. Boselli, this issue). Given the newly appreciated diversity of dwarf appearances we see that as an advantage rather than a shortcoming.

For future studies, the following procedure seems prom\-ising: Both relations seem to be related in one or the other way with the galaxies' dynamics.
\citet{2005AJ....129...61B} concluded that the CMR is a result of two more fundamental relations: 
the Faber-Jackson relation and a relation between color and velocity
dispersion. Given the slope change \linebreak of the Faber-Jackson relation from
giants to dwarfs \linebreak
\citep{2005MNRAS.362..289M,2005A&A...438..491D} a change of slope of
the CMR would actually be expected. And in the semianalytic model the sizes are strongly influenced by the dynamic feedback. In the model's context this should be closely related to the internal dynamics of the galaxies.  Therefore it is desirable to obtain central velocity dispersions and kinematics for early-type dwarfs in greater numbers.

\acknowledgements
We thank the organizers for financial support to participate in the conference. J.J.\  is supported by the Gottlieb Daimler and Karl Benz Foundation. 
   J.J.\ and T.L.\ are supported by the Excellence Initiative
    within the German Research Foundation (DFG) through the Heidelberg 
 Graduate School of Fundamental Physics (grant number GSC 129/1).
The study is based on SDSS (http://www.sdss.org/).

\clearpage

\end{document}